\documentclass[11pt]{article}
\usepackage{epsfig,graphicx}
\usepackage{amsmath,amsfonts,amssymb} 
\usepackage{color}
\usepackage{cite}
\usepackage{slashed}
\usepackage{hyperref}
\usepackage[normalem]{ulem}
\newcommand{\unit}{\leavevmode\hbox{\small1\kern-3.6pt\normalsize1}}

\def\tanb{\tan\beta}

\def\al{{A_\lambda}}
\def\ak{{A_\kappa}}

\def\ln{{\lambda_N}}
\def\lnij{{(\lambda_N)_{ij}}}
\def\aln{{A_{\lambda_N}}}
\def\alnij{{(A_{\lambda_N})_{ij}}}
\def\mn{{m^2_{\tilde{N}}}}
\def\mnij{{(m^2_{\tilde{N}})_{ij}}}
\def\ayn{A_{y_N}}
\def\aynij{(A_{y_N})_{ij}}
\def\yn{y_N}
\def\ynij{(y_N)_{ij}}

\def\lsim{\raise0.3ex\hbox{$\;<$\kern-0.75em\raise-1.1ex\hbox{$\sim\;$}}}
\def\gsim{\raise0.3ex\hbox{$\;>$\kern-0.75em\raise-1.1ex\hbox{$\sim\;$}}}

\allowdisplaybreaks

\definecolor{vmlorange}{rgb}{1.0, 0.49, 0.0}

\definecolor{vdrgreen}{rgb}{0.0, 0.7, 0.0}

\begin{document}

\thispagestyle{empty}
\begin{flushright}
  IPPP/17/56\\
  DCTP/17/112\\
  IFT-UAM/CSIC-17-064\\
  BONN-TH-2017-06\\
  EPHOU-17-011 \\
  UMN-TH-3631/17, FTPI-MINN-17/13

\vspace*{2.mm} \today
\end{flushright}

\begin{center}
  {\Large \textbf{The Constrained NMSSM with right-handed neutrinos
  } }  
  
  \vspace{0.5cm}
  David G.~Cerde\~no${}^{1,2}$,
  Valentina De Romeri${}^{3}$,
  V\'ictor Mart\'in-Lozano${}^{4}$,\\
  Keith A. Olive${}^5$,
  Osamu Seto${}^{6,7}$ \\[0.2cm] 
    
{\small \textit{ 
      ${}^1$ 
      Institute for Particle Physics Phenomenology , Department of Physics, Durham University, Durham DH1 3LE, United Kingdom\\[0pt]
      ${}^2$
      Instituto de F\'{\i}sica Te\'{o}rica
      UAM/CSIC, Universidad Aut\'{o}noma de Madrid, 28049
      Madrid, Spain\\[0pt]        ${}^3$
      AHEP Group, Instituto de F\'{\i}sica Corpuscular,
      C.S.I.C./Universitat de Val\`encia,  \\
      Calle Catedr\'atico Jos\'e Beltr\'an, 2 E-46980 Paterna (Valencia), Spain\\[0pt]
      ${}^4$ 
      Bethe Center for Theoretical Physics \& Physikalisches Institut 
      der Universit\"at Bonn,\\ 
      53115, Bonn, Germany\\[0pt] 
        ${}^5$
      William I. Fine Theoretical Physics Institute, School of Physics and Astronomy,\\
      University of Minnesota, Minneapolis, MN 55455, USA \\[0pt]
     ${}^6$
      Institute for International Collaboration,
      Hokkaido University, Sapporo 060-0815, Japan\\[0pt] 
     ${}^7$
      Department of Physics, Hokkaido University, Sapporo 060-0810, Japan }}
  

\begin{abstract}
In this article, we study the renormalization group equations of the
Next-to-Minimal Supersymmetric Standard Model, and investigate 
universality conditions on the soft supersymmetry-breaking
parameters at the Grand Unification scale.  We demonstrate that the
inclusion of right-handed neutrino superfields
can have a substantial effect on the running of the soft terms
(greatly contributing to driving the singlet Higgs mass-squared
parameter negative), which makes it considerably easier to satisfy the
conditions for radiative electroweak symmetry-breaking.
The new fields also lead to larger values of the Standard Model Higgs
mass, thus making it easier to reproduce the measured value.
We investigate the phenomenology of this constrained scenario, and
focus on two viable benchmark points, which feature a neutral lightest
supersymmetric particle (either the lightest neutralino or the
right-handed sneutrino). We show that all bounds from colliders and
low-energy observables can be fulfilled in wide areas of the parameter
space. However, the relic density in these regions is generally too high requiring some form 
of late entropy production to dilute the density of the lightest supersymmetric particle.
\end{abstract}

\end{center}

\newpage


\section{Introduction}

The Next-to-Minimal Supersymmetric Standard Model (NMSSM) is a well-motivated construction that addresses the $\mu$ problem of the MSSM through the inclusion of an extra singlet field, $S$, which mixes with the Higgs $SU(2)$ doublets and whose vacuum expectation value after electroweak symmetry breaking (EWSB) generates an effective EW-scale $\mu$ parameter~\cite{Kim:1983dt} (see, e.g., Ref.~\cite{Ellwanger:2009dp} for a review).
Among its many virtues, the NMSSM possesses a very interesting phenomenology, mainly due to its enlarged Higgs sector. For example, the mixing of the Higgs doublet with the new singlet field opens the door to very light scalar and pseudoscalar Higgs bosons with interesting prospects for collider searches.
Moreover, in the NMSSM the mass of the Higgs boson also receives new tree-level contributions from the new terms in the superpotential~\cite{Cvetic:1997ky,Barger:2006dh}, which can make it easier to reproduce the observed value~\cite{Hall:2011aa,Ellwanger:2011aa,Arvanitaki:2011ck,King:2012is,Kang:2012sy,Cao:2012fz,Ellwanger:2012ke}.
In addition, the amount in fine-tuning of the model~\cite{BasteroGil:2000bw,Ellwanger:2014dfa,Kaminska:2014wia} is reduced, when compared to the MSSM.

In order to explain the smallness of neutrino masses, the NMSSM can be extended to include a see-saw mechanism,
by adding singlet superfields that incorporate right-handed (RH) neutrinos (and sneutrinos)~\cite{Kitano:1999qb,Deppisch:2008bp}. In the resulting extended scenario the lightest RH sneutrino state is a viable dark matter (DM) candidate~\cite{Cerdeno:2008ep} with interesting phenomenological properties and a mass that can be as small as a few GeV~\cite{Cerdeno:2014cda}.

Supersymmetric (SUSY) models are characterized by the soft supersymmetry-breaking terms. The MSSM can be defined in terms of scalar  masses, $m_a$, gaugino masses $M_i$, and trilinear parameters, $A_{ij}$. 
The NMSSM also contains a new set of couplings: a singlet trilinear superpotential coupling, $\kappa$, and the strength of mixing between the singlet and Higgs doublets, $\lambda$.  In addition, there are the corresponding supersymmetry breaking trilinear potential terms $\al$ and $\ak$.
These input parameters can be defined at low-energy, in which case they would enter directly in the corresponding mass matrices to compute the physical masses of particles after radiative EWSB. This {\em effective} approach does not address the origin of the soft terms and, instead, tries to be as general as possible. 
However, if SUSY models are understood as originating from supergravity theories (which in term can correspond to the low-energy limit of superstring models), the soft parameters can be defined at some high scale as a function of the moduli of the supergravity theory. In this case, the renormalization group equations (RGEs) are used to obtain the low-energy quantities and ultimately the mass spectrum \cite{ds,eghrz,fot}.

Although in principle the number of parameters is very large ($\gsim 100$), certain simplifying conditions can be imposed, which rely on the nature of the underlying supergravity (or superstring) model. A popular choice is to consider that the soft parameters are {\em universal} at the Grand Unification (GUT) scale, i.e., $m_a=m_0$, $M_i=m_{1/2}$, and $A_{ij}=A_0$ \cite{Drees:1992am,Kane:1993td,Ellis:1996xu,Ellis:1997wva,Baer:1995nc,Baer:1997ai,Ellis:2002rp,Ellis:2003cw,Chattopadhyay:2003xi,Ellis:2012aa}.
When applied to the MSSM, the resulting Constrained MSSM (CMSSM) has only four free parameters (including the ratio of the Higgs expectation values, $\tan\beta$) plus the sign of the $\mu$ parameter. The phenomenology of the CMSSM has been thoroughly investigated in the past decades. Current Large Hadron Collider (LHC) constraints set stringent lower bounds on the common scalar and gaugino masses, while viable neutralino DM further restricts the available regions of the parameter space (for an update of all these constraints,  see Ref.~\cite{Buchmueller:2013rsa,Bagnaschi:2015eha}).

The universality condition can also be imposed in the context of the NMSSM.
The resulting constrained NMSSM (CNMSSM) also contains four free parameters which we choose as\footnote{Note that in the CMSSM, the value of $\mu$
and the supersymmetry breaking bilinear term, $B_0$, are fixed by the two conditions derived in the minimization of the Higgs potential.
In the NMSSM, we lose $\mu$ and $B_0$ as free parameters (the latter is replaced with $A_\lambda$, which is set equal to $A_0$). Thus, the two additional parameters $\lambda$ and $\kappa$, can be fixed by the three minimization conditions (which must also fix the expectation
value of the scalar component of $S$).  In practice, as will be discussed in more detail below, we allow $\lambda$ to remain
free, using the minimization conditions to fix $\kappa$ and $\tan \beta$. In this sense, the CNMSSM is constructed from the {\em same}
number of free parameters as used in the CMSSM.}: $m_0$, $m_{1/2}$, $\lambda$, and $A_0=\al=\ak$, and its phenomenology has been discussed in detail in Ref.~\cite{Djouadi:2008uj}.
It was pointed out there that recovering universal conditions for the singlet mass at the GUT scale with the correct EW vacuum at the low energy often requires a small universal scalar mass, satisfying $3 m_0 \sim - A_0 \ll m_{1/2}$.
In order for the singlet Higgs field to develop a vacuum expectation value (VEV) to fix the EW vacuum, we must require that 
$|A_0|$ is large compared to $m_0$.
As a consequence, particularly due to small $m_0$, the predicted mass range of the SM-like Higgs boson is hard to reconcile with the observed value of $m_h \simeq 125$ GeV.
In addition, large $|A_0|$ (compared to $m_0$) is also problematic as in this case, the stau tends to be tachyonic. 
In fact, this is one of the main obstacles for obtaining the observed value for the Higgs boson mass.
Furthermore, in the CNMSSM, the lightest SUSY particle (LSP) is generally either the lighter stau or the singlino-like neutralino~\cite{Ellwanger:2014hia,Ellwanger:2015axj}.
The stau, being a charged particle, can not be dark matter and
the appropriate thermal relic abundance of the singlino-like neutralino can only be realized only for limited stau-neutralino co-annihilation regions. 

In this paper, we show that these problems can be alleviated if the NMSSM is extended to include RH neutrino superfields, 
which couple to the singlet Higgs through a new term in the superpotential. 
First, the extra contributions to the RGEs help achieve unification of the soft masses for 
smaller values of the scalar and gaugino masses. This also allows more flexibility in the choice of the trilinear parameters.
Due to the RGE running of the soft mass of singlet Higgs field through its couplings with RH neutrinos, the realization of the EW vacuum becomes somewhat easier than in the NMSSM without RH neutrinos. 
We find that the lightest RH sneutrino can be the LSP in wide areas of the parameter space, where the smallest coupling between RH neutrinos and the singlet Higgs field needs to be as small as $\ln\sim10^{-4}$. As a result, the stau LSP region is significantly reduced and scalar masses as large as $m_0 \sim 10^3$ GeV are possible, 
making it easier to obtain a SM-like Higgs boson with the right mass. 
Likewise, for the neutralino LSP case with moderate values of $\ln\sim10^{-2}$, the modification of the RGE of the singlet Higgs is effective and expands (reduces)
the neutralino (stau) LSP region. As the result, in this case as well, the observed SM-like Higgs boson mass can be obtained.
In both cases the small couplings to SM particles of either the RH sneutrino LSP or the neutralino LSP  result in a  thermal relic abundance which is in excess of the observed DM density and some kind of late-time dilution is needed.

The structure of this article is the following. In Section~\ref{sec:rges}, we review the main features of the NMSSM with RH sneutrinos, we study the RGEs of the Higgs parameters, comparing them to those of the usual NMSSM, and we describe our numerical procedure. In Section \ref{sec:results}, we carry out an exploration of the parameter space of the theory, including current experimental constraints, and study the viable regions with either a neutralino or RH sneutrino LSP. We also compare our results with the ordinary NMSSM.
Finally, our conclusions are presented in Section \ref{sec:conclusions}. Relevant minimization equations and beta functions are
given in the Appendix.

\section{RGEs and universality condition}
\label{sec:rges}

The NMSSM is an extension of the MSSM and includes new superpotential terms
\begin{equation}
W_{\rm{NMSSM}} = (y_u)_{ij} Q_i \cdot H_2 U_j +(y_d)_{ij} Q_i \cdot H_1 D_j + (y_e)_{ij} L_i \cdot H_1 E_j 
+ \lambda S H_1 \cdot H_2 + \kappa S^3\ ,
\label{eq:superpotential}
\end{equation}
where the dot is the antisymmetric product and flavour indices, $i,j=1,\,2,\,3$, are explicitly included.
The model discussed here consists of the full NMSSM, 
and is extended by adding RH neutrino/sneutrino chiral superfields. This model was introduced in Refs.~\cite{Cerdeno:2008ep,Cerdeno:2009dv} (based on the construction in~\cite{Kitano:1999qb,Deppisch:2008bp}), where it was shown that the lightest RH sneutrino state is a viable candidate for DM. In previous works, only one RH neutrino superfield was considered, but here we extend the construction to include three families, $N_i$, in analogy with the rest of the SM fields and to account for three massive active neutrinos.
The NMSSM superpotential, $W_{\rm{NMSSM}}$, has to be extended in order to accommodate these new states,
\begin{equation}
W=W_{\rm{NMSSM}}+\lnij S N_i N_j + \ynij L_i \cdot H_2 N_j\ .
\label{eq:superpotential}
\end{equation}
 The new terms link the new chiral superfields with the singlet Higgs, $S$, with couplings $\ln$. Similarly, the new Yukawa interactions, $\yn$, couple the RH neutrino superfields to the second doublet Higgs, $H_2$, and the lepton doublet, $L$. 
In addition, the total Lagrangian of the model is, 
\begin{equation}
-\mathcal{L}=-\mathcal{L}_{\rm{NMSSM}} + \mnij \tilde{N}_i\tilde{N}_j^* + \left(\lnij \alnij S\tilde{N}_i\tilde{N}_j  +\ynij\aynij\bar{L}_iH_2 \tilde{N}_j + h.c. \right),
\label{eq:softlagrangian}
\end{equation}
where $\mathcal{L}_{NMSSM}$ includes the scalar mass terms and trilinears terms of the NMSSM and $\mathcal{L}$
includes new $3\times 3$ matrices of trilinear parameters, $\aln$ and $\ayn$, and a $3\times 3$  matrix of squared soft masses for the RH sneutrino fields, $\mn$. In our analysis, we will consider that all these matrices are diagonal at the GUT scale. 
As pointed out in Ref.~\cite{Cerdeno:2008ep}, the neutrino Yukawa parameters are small, $\ynij\lesssim 10^{-6}$, since the neutrino Majorana masses generated after EWSB are naturally of the order of the EW scale. Thus, they play no relevant role in the RGEs of the model and can be safely neglected.
The new parameters ($\lambda_N, A_{\lambda_N}$) are chosen to be real.
Finally, we will extend the universality conditions to the new soft parameters, thus demanding 
 \begin{eqnarray} 
 m_S^2&=&m_0^2\,,\nonumber\\
\mnij&=&diag\left( m^2_0,\, m^2_0,\,m^2_0\right)\, ,\nonumber\\
\lnij&=&diag\left( \lambda_{N_1},\, \lambda_{N_2},\,\lambda_{N_3} \right)\ ,\nonumber\\
\al=\ak&=&A_0 \, ,\nonumber \\
\alnij=\aynij&=&diag\left( A_0,\, A_0,\,A_0\right)\ ,
\label{eq:unificationcond}
\end{eqnarray}
 at the GUT scale, which is defined as the scale where gauge couplings of $SU(2)_L$ and $U(1)_Y$ coincide.

\subsection{Radiative EW symmetry-breaking and the singlet soft mass}

Using the values of the soft terms, defined at the GUT scale, the RGEs can be numerically integrated down to the EW scale. After EWSB, the minimization conditions of the scalar potential leave three tadpole equations for the VEVs of the three Higgs fields. At tree level, these are
\begin{eqnarray}
\frac{\partial V}{\partial \phi_d}&=&\frac{v_sv_u \lambda}{2} (- \sqrt{2}A_\lambda - \kappa v_s)-\frac{(g_1^2+g_2^2)}{8}v_d(v_u^2-v_d^2)+ m_{H_d}^2v_d + \frac{\lambda^2}{2}(v_s^2+v_u^2)v_d, \label{Eq:Stat:vd}\\
\frac{\partial V}{\partial \phi_u}&=&\frac{v_sv_d \lambda}{2} (- \sqrt{2}A_\lambda - \kappa v_s)+\frac{(g_1^2+g_2^2)}{8}v_u(v_u^2-v_d^2)+ m_{H_u}^2v_u + \frac{\lambda^2}{2}(v_s^2+v_d^2)v_u, \label{Eq:Stat:vu}\\
\frac{\partial V}{\partial \phi_s}&=&\frac{v_s}{2}(\sqrt{2}A_\kappa \kappa v_s + 2m_S^2 + \lambda^2(v_d^2+v_u^2)-2\kappa\lambda v_u v_d+2\kappa^2v_s^2)-\frac{A_\lambda\lambda}{\sqrt{2}} v_uv_d.\label{Eq:Stat:vs}
\end{eqnarray}
As noted earlier, using the measured value of the mass of the $Z$ boson, $M_Z$, and its relation to the Higgs doublet VEVs, $v_u$ and $v_d$, the conditions for correct EWSB allow us to determine the combination $\tan\beta\equiv v_u/v_d$, and $v_s$, as well as one additional parameter which we take as $\kappa$.
Thus, the constrained version of the NMSSM can be defined in terms of four universal input parameters, 
\begin{equation}
m_0,\ m_{1/2},\ \lambda ,\ A_0=\al=\ak\,.
\label{Eq:inputs}
\end{equation}
In practice, however, solving the system of tadpole equations is in general easier if one fixes the value of $\tan \beta$ and uses the 
tadpole conditions to determine the soft mass of the singlet Higgs, $m_S^2$. Although this generally results in a non-universal mass for $m_S$,  it is then possible to iteratively find the value of $\tan \beta$ such that $m_S = m_0$. 

More specifically, using the above tree-level expressions (for illustrative purposes), a combination of Eqs.~\eqref{Eq:Stat:vd} and \eqref{Eq:Stat:vu} leads to
\begin{eqnarray}
 \mu_{\rm eff}^2 \equiv \frac{1}{2} (\lambda v_s)^2 =
  -\frac{1}{2}M_Z^2
  -  \frac{ m_{H_u}^2 \tan\beta^2 -m_{H_d}^2 }{\tan\beta^2 - 1}.
\label{Setlow:mueff}
\end{eqnarray}
Since $\lambda$ is an input free parameter, we can use it to define $v_s$ as 
\begin{equation}
 v_s = \pm \sqrt{\frac{2 \mu_{\rm eff}^2}{\lambda^2}}.
 \label{eq:tadvs}
\end{equation}
The sign of $v_s$ plays the role of the sign of $\mu$-term in the CMSSM.
From another combination of Eqs.~\eqref{Eq:Stat:vd} and \eqref{Eq:Stat:vu} we obtain
\begin{equation}
 (B\mu)_{\rm eff} \equiv 
 \frac{\lambda v_s}{\sqrt{2}} ( A_{\lambda} + \frac{1}{\sqrt{2}} \kappa v_s)
 =  \frac{\sin2\beta}{2}( m_{H_u}^2 +m_{H_d}^2 + 2\mu_{\rm eff}^2 ),
\label{stateq:doublet}
\end{equation}
which allows us to solve for $\kappa$
\begin{equation}
\kappa
 = \frac{\sqrt{2}}{v_s}\left( -A_{\lambda} + \frac{(B\mu)_{\rm eff} }{sgn(\mu_{\rm eff})\mu_{\rm eff}} \right) .
 \label{eq:tadkappa}
\end{equation}
For the last parameter, $m_S^2$, we can use Eq.~\eqref{Eq:Stat:vs} in the form of 
\begin{equation}
 m_S^2 
 = - \left( \frac{1}{\sqrt{2}}A_{\kappa} \kappa v_s +\frac{1}{2} \lambda^2 (v_d^2 + v_u^2)
  - \kappa \lambda v_u v_d + \kappa^2 v_s^2 \right) + \frac{1}{\sqrt{2}v_s}A_{\lambda} \lambda v_u v_d.
  \label{eq:tadms2}
\end{equation}
The one-loop expressions can be found in the Appendix \ref{app:loopedminimization}.
The above procedure assumes $\tan \beta$ is free, but in our analysis we add one extra step: for each point in the parameter space, we vary the value of $\tan\beta$ in order to impose $m_S^2({\rm GUT})=m_0^2$ (within a certain tolerance ($\sim 1$ \%)). If this universality condition cannot be achieved, the point is discarded. This procedure was outlined in Ref.~\cite{Djouadi:2008yj}. Thus, at the end of this iterative process, the free parameters are those in Eq.~\eqref{Eq:inputs}.

This prescription has been applied in the literature to study the phenomenology of the CNMSSM. A first thing to point out is that the resulting value of $m_S^2$ at the EW scale from Eq.~\eqref{eq:tadms2} is often negative~\cite{Ellwanger:2006rn}, and this makes it difficult to satisfy the universality condition. 
In particular, it was found in~\cite{Djouadi:2008uj}, that the resulting value of $\tan\beta$ in the CNMSSM is in general large and that, in general, the value of the universal gaugino mass is also large. As a result, the lightest stau is the LSP in the remaining viable areas of the parameter space (which poses a problem to incorporate DM in this scenario).  
In order to alleviate this, a semi-constrained version of the NMSSM was explored in Ref.~\cite{Ellwanger:2006rn}, allowing for $m_S^2\neq m_0^2$ and $A_\kappa\neq A_0$ at the GUT sale.

In our extended model, the solution of the tadpole equations proceeds in the same way as in the CNMSSM. However, as we will argue in Section~\ref{sec:results}, the RH sneutrino contributes to the RGEs of the singlet and singlino and opens up the parameter space allowing
us to restore full universality. 

In particular, the new terms in the superpotential and the soft breaking parameters 
enter the 1-loop beta function for the scalar mass of the singlet Higgs, $m_S^2$, which is now given by
\begin{eqnarray}
\beta_{m_S^2}^{(1)} =  
4 \Big(
3 m_S^2 |\kappa|^2  + |T_{\kappa}|^2 + |T_{\lambda}|^2 +
\left(m_{H_d}^2 + m_{H_u}^2 + m_S^2\right)|\lambda|^2\notag\\  
+ m_S^2 \mbox{Tr}\left({\lambda_N  \lambda_N }\right)  
+2 \mbox{Tr}\left({\mn  \lambda_N  \lambda_N }\right)
+ \mbox{Tr}\left(T_{\lambda_N} T_{\lambda_N}
\right)
\Big).
\label{eq:1loop}
\end{eqnarray}
We have defined
$T_{g_i}= A g_{i}$, where $A$ is the soft trilinear term and $g_i$ is the corresponding coupling constant, $g_i= y_i,\ \lambda ,\ \kappa,\ \lambda_N$.
The first line corresponds to the usual NMSSM result, and the second line contains the new contribution from the coupling of the singlet to 
the right-handed neutrino. For completeness, the two-loop expression is given in Eq.~(\ref{eq:twoloop}). 

\begin{figure}
      \begin{center}
\scalebox{0.8}{
      \includegraphics[scale=0.5]{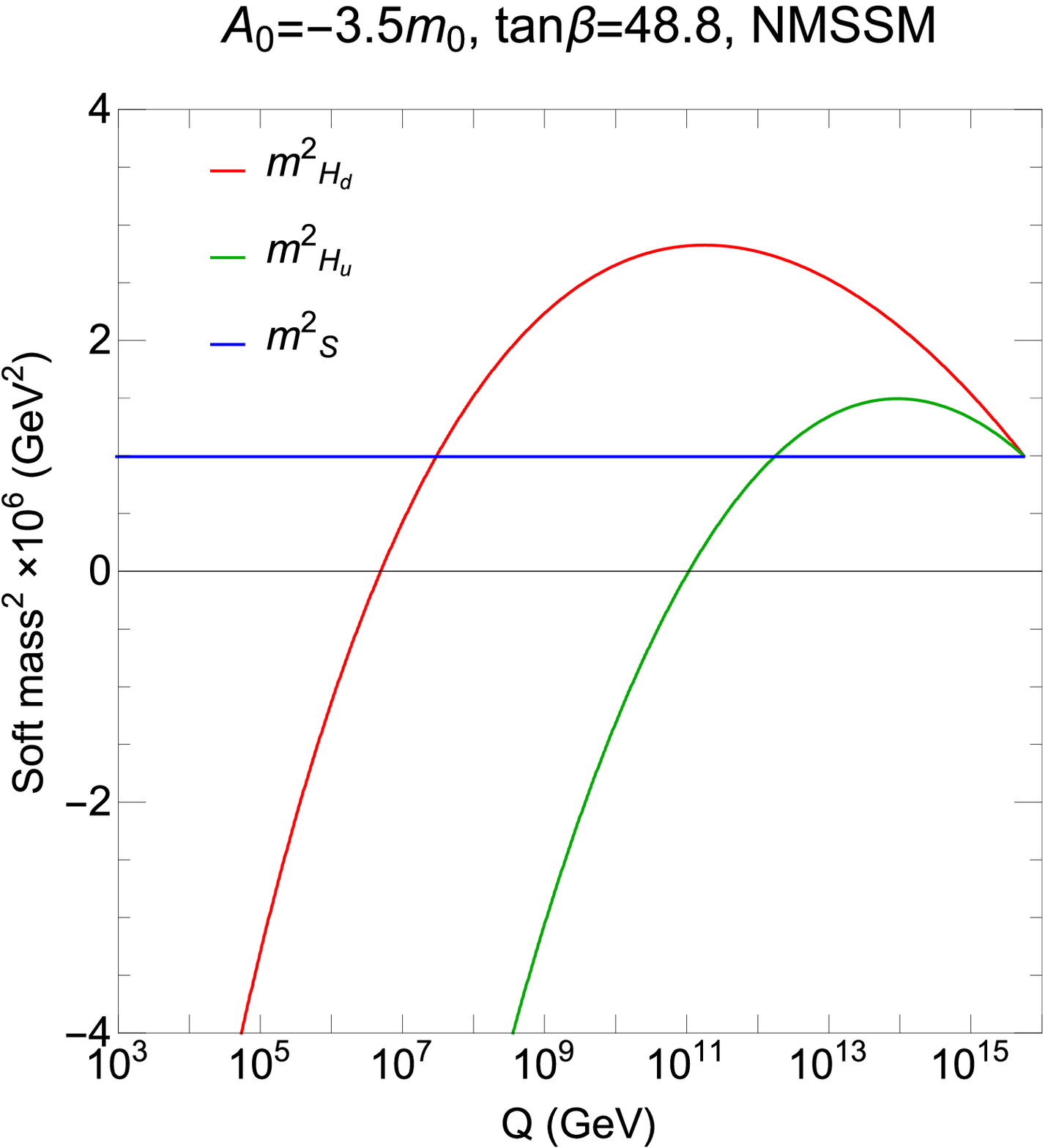}
            \includegraphics[scale=0.5]{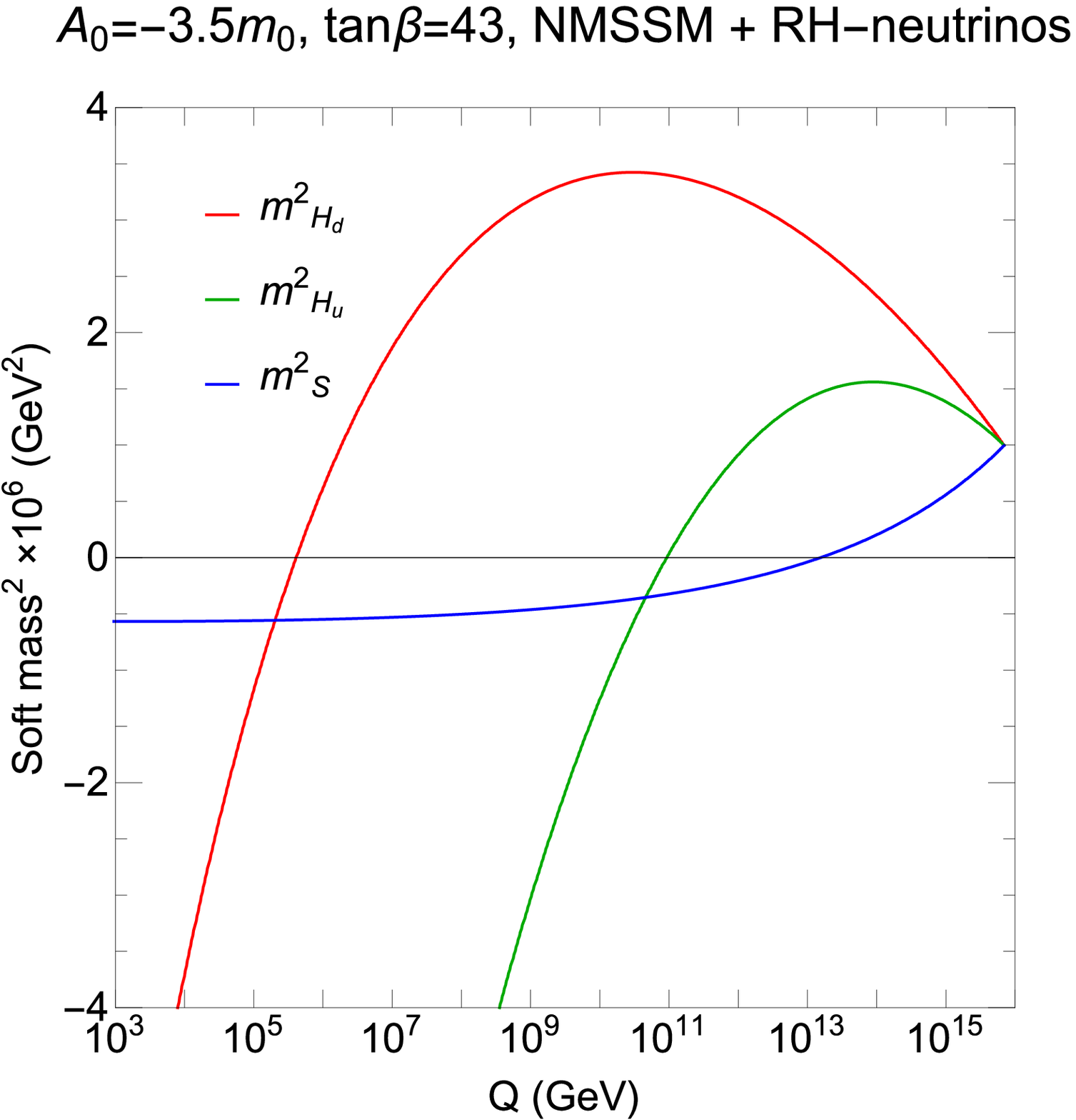}
               }
      \caption{\footnotesize
 2-loop RGE running of the soft Higgs mass parameters, $m_{H_d}^2$, $m_{H_u}^2$, and $m_S^2$, imposing the universality condition $m_0=1000$~GeV at the GUT scale, 
 with $A_0=-3.5\,m_0$,  $m_{1/2}=4500$~GeV, and $\lambda = 0.01$ (the latter is input at the weak scale). 
 The plot on the left corresponds to the standard NMSSM (i.e., with $\ln=0$). The plot on the right corresponds to the extended NMSSM with RH neutrinos for 
 $\lambda_N=(0.0002, 0.6,0.6)$, defined at the GUT scale. 
 The value of $\tan\beta$ has been fixed separately in each example in order to achieve universality. }
      \label{fig:mv2tln}
      \end{center}
\end{figure}

We show in Fig.\,\ref{fig:mv2tln} the running of the Higgs mass-squared parameters as a function of the renormalization scale. We have chosen an example where the soft terms unify at the GUT scale in the standard NMSSM (left) and in the extended NMSSM with RH neutrinos (right). As the RGE running in the two models differs, we require slightly different values of $\tan \beta$ to achieve $m_S = m_0$. 
Enforcing the unification of the scalar singlet mass tends to be problematic for radiative EWSB
in models without the right-handed neutrino, as $m_S^2$ remains positive down to the weak scale. 
As we can observe, the effect of the RH sneutrino fields in the running of the $m_S^2$ parameter is remarkable. 
In this example, it can drive the positive singlet mass-squared term negative. This alleviates some tension in the choice of initial parameters.

\subsection{Details on the numerical code}
\label{subsec:numcode}

We have modified the supersymmetric spectrum calculator  {\tt SSARD} \cite{ssard} by adding the necessary RGEs to include additional
terms needed in our extension of the NMSSM.  The code numerically integrates the RGEs between the weak and GUT scales and solves the tadpole equations used to determine $\kappa$, $v_s$ and $m^2_S$ as outlined above. The output of this program is then passed through the public packages {\tt NMSSMTools 4.9.2}~\cite{Ellwanger:2004xm,Ellwanger:2005dv,Ellwanger:2006rn} and {\tt Micromegas 4.3}~\cite{Belanger:2014vza} in order to get the physical particle spectrum and the thermal component to the DM relic abundance.

{\tt SSARD} implements an iterative procedure to solve the RGEs as follows. Using weak scale inputs for the gauge and Yukawa couplings, the GUT scale is defined as the renormalization scale where the $SU(2)_L$ and $U(1)_Y$ gauge couplings coincide. 
At this GUT scale, universal boundary conditions are imposed for all gaugino masses, $m_{1/2}$,
trilinear terms, $A_i=A_\lambda=A_\kappa=A_0$,
and scalar masses, $m^2_i=m_0^2$, but we leave $m_S^2({\rm GUT})$ as a free parameter. 
The couplings $\lambda_N$ are also input at the GUT scale. 
We then run the RGEs from the GUT to the SUSY scale, where we solve the tadpole equations (now including the tadpole condition for $S$) with the resulting values of the parameters. The coupling $\lambda$ is input at the weak scale. 
Using these low-scale values, we then run the RGEs upwards, recalculating the GUT scale, and we iterate this procedure until a good stable solution is found. 
As a final step, this procedure is repeated for different values of $\tan\beta$, searching for points in which the unification condition 
$|1 - m_S^2({\rm GUT})/m_0^2| < 10^{-2}$ is satisfied.

Once the tadpole equations are solved for the points that fulfill the universality conditions, we collect all the parameters at EW scale and compute the 
SUSY spectrum using the public package {\tt NMSSMTools 4.9.2} ~\cite{Ellwanger:2004xm,Ellwanger:2005dv,Ellwanger:2006rn}. 
The code checks the scalar potential, looking for tachyonic states, the correct EW vacuum, divergences of the coupling at some scale 
between the SUSY and GUT scales, as well as collider constraints from LEP and LHC, and low energy observables. 
If a point is allowed, the program computes the SUSY spectrum for the given set of parameter values as well as the SM-like Higgs mass 
with full 1-loop contributions and the 2-loop corrections from the top and bottom Yukawa couplings.

In order to test our procedure, we have also implemented our model in  {\tt SARAH}~\cite{Staub:2009bi,Staub:2010jh,Staub:2012pb,Staub:2013tta,Staub:2015kfa}, 
which produces the model files for {\tt SPheno}~\cite{Porod:2003um,Porod:2011nf} to perform the running from the GUT to the EW scale.
We notice that even a ``small" variation (within 10\%) of the parameters given as input to the numerical codes (such as $\lambda$, $A_0$, $m_0$, $m_{1/2}$) 
can lead to very different values of the outputs - in particular of
$A_\lambda, \kappa$ and $m_S^2$. On the other hand, $v_s$ turns out not to be affected much by these variations, 
since its tadpole equation depends mostly on $\rm tan \beta$, when $\rm tan \beta$ is large.
In particular, $A_\lambda$ is the most numerically unstable parameter. 
This instability may induce differences in the soft mass of the singlet Higgs $m_S^2$, although its RGE is rather stable and its low-scale value is only affected through the stationary conditions. Eventually, $\rm tan \beta$ is the most sensitive parameter to change outputs significantly. However its value is finally fixed by imposing the universality condition $m_S^2 = m_0^2$ and therefore all the eventual differences in the parameters get reabsorbed. We have carried out several tests and we have found an agreement within a 10\% between both codes. Moreover, we have also tested the codes in the pure NMSSM limit and we have found an agreement within a 10\% between {\tt SSARD} and {\tt NMSSMTools}.

\section{Results}
\label{sec:results}

In this section, we provide some numerical examples that illustrate the effect of adding RH sneutrinos in the four-dimensional NMSSM parameter space with universal conditions. Rather than performing a full numerical scan on all the parameters, we have selected some representative ($m_{1/2},m_0$) slices, and fixed $\lambda=0.01$,  $A_0=-3.5\,m_0$.
The condition $3 m_0 \sim - A_0 \ll m_{1/2}$ is required to get the correct EW vacuum~\cite{Djouadi:2008uj}, as already stated in the Introduction.
In agreement with observed values, we have also fixed $m_{\rm top} =173.2$ GeV, $m_{\rm bottom} = 4.2$ GeV. 

We have investigated three different scenarios. First, for comparison, we consider the Constrained NMSSM case, and then we study two scenarios of the extended model with RH sneutrinos. In particular, we consider one scenario with $\lambda_N=(0.0002, 0.6,0.6)$ (``\textit{small} $\lambda_N$'') and another one with $\lambda_N=(0.01, 0.6,0.6)$ (``\textit{large} $\lambda_N$''). The ``\textit{small} $\lambda_N$'' scenario is motivated by the fact that the RH sneutrino can be the LSP whereas in the ``\textit{large} $\lambda_N$'' the lightest neutralino can be the LSP.

\begin{figure}[t]
	\begin{center}
		\includegraphics[scale=0.45]{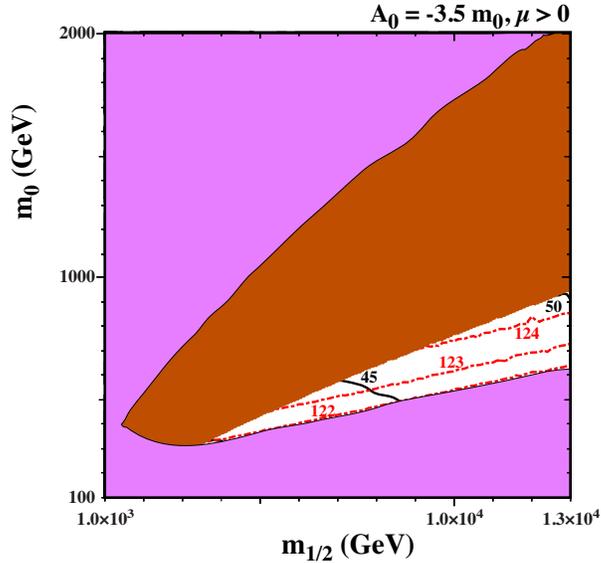}
		\caption{\footnotesize
			Higgs mass contour plot in the plane ($m_{1/2}$-$m_0$) for the CNMSSM scenario. We depict in magenta the region of the parameter space excluded by any of the following reasons: existence of another vacuum deeper than the EW one; the presence of a tachyonic particle; experimental constraints from LEP, LHC and others (see text for a detailed description). In the brown shaded area the stau is the LSP while in the white area the neutralino is the LSP. Red dashed contours account for the Higgs mass (in GeV), while the black lines represent the value of $\tanb$.}
		\label{fig:nmssmcase}
	\end{center}
\end{figure}

\paragraph{CNMSSM:}

Let us first focus on the pure CNMSSM case without RH neutrino fields. In Fig.~\ref{fig:nmssmcase}, we show the results of a numerical scan in the plane ($m_{1/2},m_0$).
We have imposed consistency with all experimental results, including ATLAS scalar searches~\cite{ATLAS:2014vga}, bounds on low energy observables, 
such as $B_s\to \mu^+\mu^-$~\cite{Bobeth:2013uxa,Domingo:2015wyn} and $b \to s +\gamma$~\cite{Misiak:2015xwa,Domingo:2015wyn} by {\tt NMSSMTools}, 
and collider constraints on the masses of SUSY particles. 
In Fig.~\ref{fig:nmssmcase}, the magenta area for large $m_0$ corresponds to parameter values which lead to a tachyonic stau, 
whereas for small $m_0$ it is due to the ATLAS $h^0/H^0/A^0 \rightarrow\gamma\gamma$ searches~\cite{ATLAS:2014vga}, 
which can be used as a constraint on searches of a light Higgs boson that often appears in the general NMSSM 
(this essentially rules out the region of the parameter space with $m_h < 122$ GeV).
Since the purpose of this paper is not to explain anomalies such as those observed in the measurement of the muon anomalous magnetic moment, $(g-2)_\mu$, or the $B^+ \to \tau^+ \nu_{\tau}$ branching ratio, we do not restrict our interest to such a parameter region. 
The magenta area also represents an unavailable or excluded region where either the universal conditions are not realized, there are deeper vacua than the EW one, a sfermion or any Higgs boson is tachyonic, or any experimental bound is not fulfilled according to the constraints described in Section~\ref{subsec:numcode}.

The brown shaded area corresponds to the solutions where the universal conditions are fulfilled but the stau is the LSP, whereas in the remaining white area, the neutralino is the LSP. 
The black contours represent the values of $\tanb$ necessary to achieve the universal conditions (seen here to lie in the range of $\tanb\sim40 - 50$), while the red dot-dashed contours show the SM-like Higgs mass. We notice that the experimentally observed Higgs mass is not achieved in the allowed region. Indeed, the highest value for the SM-like Higgs mass is around 124 GeV for large values of $\tanb$ ($\sim 50$), although this region 
remains acceptable if we consider a $\pm 2$ GeV uncertainty in the calculation of the Higgs mass.
It has been pointed out in Ref.~\cite{Djouadi:2008uj} that the stau-neutralino coannihilation strip in the CNMSSM extends only up to values of $m_{1/2}$ of the order of a few TeV, which roughly corresponds to $m_{\tilde{\tau}_1} \lesssim 1$ TeV. 
In this plot, this region is excluded due to constraints in the Higgs sector, as explained above.

\begin{figure}[t]
	\begin{center}
		\includegraphics[scale=0.45]{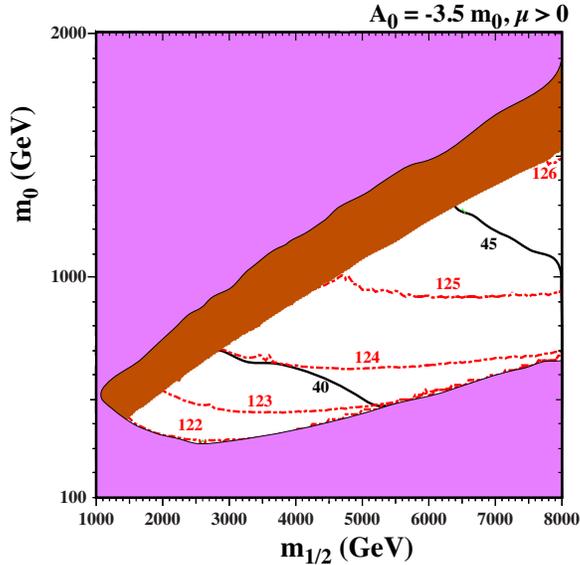}
		\caption{\footnotesize
			Higgs mass contour plot in the plane ($m_{1/2}$-$m_0$) for the ``\textit{small} $\lambda_N$'' scenario with $\lambda_N=(0.0002,0.6,0.6)$. The colour code is the same as in Fig.~\ref{fig:nmssmcase}, except in the white region that represents the case where the sneutrino is the LSP in this case. Red dashed contours account for the Higgs mass (in GeV), while the black lines represent the value of $\tanb$.}
		\label{fig:smalllambda}
	\end{center}
\end{figure}

\paragraph{Small $\ln$ scenario:}

Next, we concentrate on our extended model, when the RH sneutrino field is added to the particle content of the NMSSM. In Fig.~\ref{fig:smalllambda}, we show the results of a scan in the ($m_{1/2},m_0$) plane, 
for the ``\textit{small} $\lambda_N$'' scenario, $\lambda_N=(0.0002,0.6,0.6)$. The colour code in this figure is the same as in Fig.~\ref{fig:nmssmcase}. 
The excluded magenta areas are due to tachyonic staus (for large $m_0$), tachyonic RH sneutrino (for a portion of small $m_0$ and large $m_{1/2}$), and due to the ATLAS bound on  $h^0/H^0/A^0 \rightarrow \gamma\gamma$ (for the small $m_0$ region).
The allowed parameter space differs from that obtained in the CNMSSM. In particular, greater values of $m_0$ are allowed. Interestingly, this leads to larger values of the Higgs mass and the correct value ($\sim 125$ GeV) can be achieved for $0.9 \lesssim m_0\lesssim 1$ TeV, $m_{1/2}\gtrsim 4.5$ TeV and $\tan \beta \gtrsim 40$. 
In the allowed area of this scenario, the RH sneutrino is the LSP. 
Since the RH neutrino Majorana mass term is proportional to $\ln$, and this is also the leading contribution to the RH sneutrino mass, 
small values $\ln \sim 10^{-4}$, are favoured to a obtain RH sneutrino LSP. 
Notice however that for such a small value of the coupling, the annihilation rate of the RH sneutrino into SM particles is in general very small 
and the resulting thermal relic density is too large. Thus, the viability of this model would entail some sort of dilution mechanism at late times.

\begin{figure}[t]
	\begin{center}
		\includegraphics[scale=0.45]{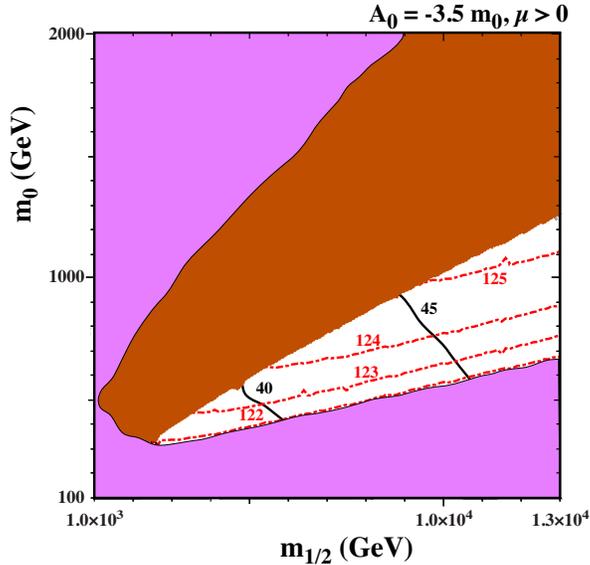}
		\caption{\footnotesize
			Higgs mass contour plot in the plane ($m_{1/2}$-$m_0$) for the ``\textit{large} $\lambda_N$'' scenario with $\lambda_N=(0.01,0.6,0.6)$. The colour code is the same as in Fig.~\ref{fig:nmssmcase}. In this scenario a light neutralino is the LSP in the white areas. Red dashed contours account for the Higgs mass (in GeV), while the black lines represent the value of $\tanb$.}
		\label{fig:biglambda}
	\end{center}
\end{figure}

\paragraph{Large $\ln$ scenario:}

An interesting alternative is to work in the ``\textit{large} $\lambda_N$'' regime. In Fig.~\ref{fig:biglambda} we show the scan result in the ($m_{1/2},m_0$), now taking $\lambda_N=(0.01,0.6,0.6)$. With a larger $\ln$, the resulting mass of the lightest RH sneutrino as well as that of the RH neutrino increase and hence the LSP is found to be either the neutralino or stau. In the allowed area of Fig.~\ref{fig:biglambda} the lightest neutralino is the LSP while the brown area shows where the stau is the LSP as in previous figures. We notice also that in this scenario a larger value of $m_{1/2} \gsim 900$ GeV is required in order to reproduce the observed Higgs mass.

As we demonstrated in the previous examples, the inclusion of RH neutrinos expands the parameter region of the neutral LSP compared with the CNMSSM case, however the difficulty of achieving the thermal relic abundance of DM is not improved. The reason is the same as in the pure CNMSSM mentioned above. The lower bound on the Higgs boson mass, $m_h>122$ GeV, sets bounds on the soft masses that are $m_{1/2} \gsim$ a few TeV and $m_{0} (m_{\tilde{\tau}_1}) \lesssim 1$ TeV, where the annihilation cross section of $\tilde{\tau}$ is smaller than about $1$ pb. Hence, even with strong coannihilation with staus, the resultant thermal relic abundance of the neutralino LSP is too large leaving $\Omega~h^2 > 0.12$. For the RH sneutrino LSP in the ``\textit{small} $\lambda_N$'' scenario, the main annihilation modes are $\tilde{N}\tilde{N} \rightarrow \rm W^+ W^-,~Z^0 Z^0, ...$ through Higgs boson exchange, with a cross section that is also suppressed by small $\ln$, ending up with a huge thermal relic abundance. One may then search for possible coannihilation effects with stau NLSP in the parameter region where $\tilde{N}$ is quasi-degenerate with $\tilde{\tau}_1$. However, unfortunately this is not the case. In addition to the fact that annihilation cross section of stau is smaller than $1$ pb for $m_{\tilde{\tau}_1} \lesssim 1$ TeV as mentioned above, the coannihilating particles $\tilde{N}$ and $\tilde{\tau}$ are actually decoupled from each other, because the reaction rates of all processes between $\tilde{N}$ and $\tilde{\tau}$ such as $\tilde{\tau},\tilde{N} \rightarrow X, Y$ and $\tilde{\tau}, X \rightarrow \tilde{N}, Y$, with $X, Y$ being possible SM particles, are negligible due to small $\lambda_N$ of the order of $10^{-4}$ with heavy mediating neutralinos. Hence, in both scenarios with ``\textit{large} $\lambda_N$'' and ``\textit{small} $\lambda_N$'', if the LSP is DM, its final abundance has to be explained by nonthermal mechanisms. However in fact, within the framework of supergravity or superstring, it is possible that our Universe has undergone nonstandard thermal history because many supergravity models predict moduli fields and hidden sector fields, which affect the evolution of the early Universe.
Scenarios of nonthermal DM production include, for example, (i) regulated thermal abundance by late time entropy production from moduli decay~\cite{Coughlan:1983ci,deCarlos:1993wie,ego}, thermal inflation~\cite{Lyth:1995hj,Lyth:1995ka,Asaka:1999xd} or defect decay~\cite{Kawasaki:2004rx,Hattori:2015xla}, (ii) generated by the decay of late decaying objects such as moduli~\cite{Moroi:1994rs,Kawasaki:1995cy,ego} or $Q$-balls~\cite{Enqvist:1998en}, and (iii) nonthermal scatterings and decays as studied in Refs.~\cite{McDonald:2001vt,Asaka:2005cn,Asaka:2006fs}. 

In the results of the analysis performed in this model and shown in Figs.~\ref{fig:nmssmcase},~\ref{fig:smalllambda} and~\ref{fig:biglambda} we have fixed the trilinear term $A_0=-3.5 ~m_0$. We have numerically checked the effect of changing this relation. We found that a smaller ratio $-A_0/m_0$ would require larger values of $m_0$, $m_{1/2}$ and $\tan \beta$ to reproduce the observed Higgs mass. For instance, in the scenario with ``\textit{small} $\lambda_N$'', if $A_0=-2.6\,m_0$ the  Higgs mass ($\sim 125$ GeV) is obtained for  $m_0 \sim 1.5$ TeV, $m_{1/2} \sim 6 - 8$ TeV and $\tan \beta \gtrsim 47$. 
A larger value of  $-A_0/m_0$ ratio, generally leads to Landau poles in the RGEs (as the value of $\tan\beta$ needed to obtain $m_S({\rm GUT})=m_0$ becomes too large). Finally, for the opposite sign of the trilinear parameter, $A_0$, the correct EW vacuum cannot be realized and tachyons in the Higgs sector appear.

\section{Conclusions}
\label{sec:conclusions}

In this paper we have studied an extended version of the NMSSM in which RH neutrino superfields are included through a coupling with the singlet Higgs. 
We have observed that the contributions of the new terms to the RGEs make it possible to impose universality conditions on the soft parameters, 
thus considerably opening up the parameter space of the constrained NMSSM.

We have computed the two-loop RGEs of this model and solved them numerically, using the spectrum calculator  {\tt SSARD}. The RH sneutrino coupling to the singlet Higgs leads to a contribution to the RGE of the singlet Higgs mass-squared parameter that helps driving it negative, thus making it easier to satisfy the conditions for EWSB, while imposing universality conditions at the GUT scale. This significantly alleviates the tension in the choice of initial parameters and opens up the parameter space considerably. Moreover, the RH sneutrino contribution also leads to slightly larger values of the resulting SM Higgs mass, which further eases finding viable regions of the parameter space.

We have studied two possible benchmark scenarios in which the LSP is neutral: either the lightest RH sneutrino or the lightest neutralino. In these examples, we have implemented all the recent experimental constraints on the masses of SUSY particles and on low-energy observables. Finally, we have also computed the resulting thermal dark matter relic density, but we have not imposed any constraint on this quantity.

The RH sneutrino can be the LSP, but only when its coupling to the singlet Higgs is very small ($\ln\sim10^{-4}$). This leads to very large values of the thermal relic abundance. Although there are regions in which the stau NLSP is very close in mass, coannihilation effects are negligible (since the RH sneutrino-stau annihilation diagrams are also suppressed by $\ln$.)
On the other hand, for large values of $\ln\sim10^{-2}$, the lightest neutralino can be the LSP. The remaining areas feature in general smaller values of the soft scalar mass than in the NMSSM, however, the neutralino relic abundance is also too large
requiring some form of late time dilution.

\vspace*{1cm}
\noindent{\bf \large Acknowledgments}

We are thankful to F. Staub for his help with SARAH. DGC is supported by the STFC and the partial support of the Centro de Excelencia Severo Ochoa Program through the IFT-UAM/CSIC Associate programme. VDR acknowledges support by the Spanish grant SEV-2014-0398 (MINECO) and partial support by the Spanish grants FPA2014-58183-P and PROMETEOII/2014/084 (Generalitat Valenciana).
VML acknowledges the support of the BMBF under project 05H15PDCAA.
The work of K.A.O. was supported in part by
DOE grant DE-SC0011842 at the University of Minnesota. 
We also acknowledge support by
the Consolider-Ingenio 2010 programme under grant MULTIDARK CSD2009-00064 and
the European Union under the ERC Advanced Grant SPLE under contract ERC-2012-ADG-20120216-320421.

\appendix
\section{One-loop corrected minimization equations}
\label{app:loopedminimization}

In our calculation, we have imposed the minimization condition to the effective potential 
$V = V^{\rm tree} + \Delta V^{\rm one-loop}$, including one loop corrections $\Delta V^{\rm one-loop}$.
We have three tadpole equations for one loop effective potential $V$, namely
\begin{eqnarray}
\frac{\partial V}{\partial \phi_d}&=& 0, \label{App:Eq:Stat:vd} \\
\frac{\partial V}{\partial \phi_u}&=& 0, \label{App:Eq:Stat:vu} \\
\frac{\partial V}{\partial \phi_s}&=& 0. \label{App:Eq:Stat:vs}
\end{eqnarray}
One combination of Eqs.~(\ref{App:Eq:Stat:vd}) and (\ref{App:Eq:Stat:vu}) gives a formula of the effective $\mu$ parameter as
\begin{eqnarray}
\mu_{\rm eff}^2 &\equiv& \frac{1}{2} (\lambda v_s)^2 \nonumber \\
&=& \frac{
  -\frac{1}{2}M_Z^2(\tan\beta^2 - 1)
  -  m_{H_u}^2 \tan\beta^2 + m_{H_d}^2 +\Delta_1 }{\tan\beta^2 - 1 +\Delta_2}, \\
\Delta_1 &=& -\frac{3 y_t^2}{16\pi^2} 
 \tan\beta^2\left(\mathcal{F}(m_{\tilde{t}_1},Q) + \mathcal{F}(m_{\tilde{t}_2}, Q) - 2\mathcal{F}(m_t, Q) 
 - A_t^2\frac{ \mathcal{F}(m_{\tilde{t}_1},Q) - \mathcal{F}(m_{\tilde{t}_2},Q)}{m_{\tilde{t}_2}^2-m_{\tilde{t}_1}^2}\right)  \nonumber \\
&& + \frac{3 y_b^2}{16\pi^2} 
 \left(\mathcal{F}(m_{\tilde{b}_1},Q) + \mathcal{F}(m_{\tilde{b}_2}, Q) - 2\mathcal{F}(m_b, Q)
 - A_b^2\frac{\mathcal{F}(m_{\tilde{b}_1},Q) - \mathcal{F}(m_{\tilde{b}_2},Q)}{m_{\tilde{b}_2}^2-m_{\tilde{b}_1}^2} \right)  \nonumber \\
&& +\frac{ y_{\tau}^2}{16\pi^2}
 \left(\mathcal{F}(m_{\tilde{\tau}_1},Q) + \mathcal{F}(m_{\tilde{\tau}_2}, Q) - 2\mathcal{F}(m_\tau, Q) 
 - A_{\tau}^2\frac{\mathcal{F}(m_{\tilde{\tau}_1},Q) - \mathcal{F}(m_{\tilde{\tau}_2},Q)}{m_{\tilde{\tau}_2}^2-m_{\tilde{\tau}_1}^2} \right) , \\
\Delta_2 &=& \frac{3 y_t^2 }{16\pi^2}
\frac{ \mathcal{F}(m_{\tilde{t}_1},Q) - \mathcal{F}(m_{\tilde{t}_2},Q) }{m_{\tilde{t}_2}^2-m_{\tilde{t}_1}^2} - \frac{3 y_b^2 }{16\pi^2}
 \tan\beta^2\frac{  \mathcal{F}(m_{\tilde{b}_1},Q) - \mathcal{F}(m_{\tilde{b}_2},Q)}{m_{\tilde{b}_2}^2-m_{\tilde{b}_1}^2} \nonumber \\
&& - \frac{ y_{\tau}^2}{16\pi^2}
 \tan\beta^2 \frac{  \mathcal{F}(m_{\tilde{\tau}_1},Q) - \mathcal{F}(m_{\tilde{\tau}_2},Q)}{m_{\tilde{\tau}_2}^2-m_{\tilde{\tau}_1}^2} .
\end{eqnarray}
Here, 
\begin{eqnarray}
      \mathcal{F}(m,Q) = m^2 \left(\log\left(\frac{m^2 }{Q^2}\right) -1 \right) ,
\end{eqnarray}
is an auxiliary function.

Another combination of Eqs.~(\ref{App:Eq:Stat:vd}) and (\ref{App:Eq:Stat:vu}) gives a formula of the effective $B \mu$ as
\begin{eqnarray}
(B\mu)_{\rm eff} &\equiv& \frac{\lambda v_s}{\sqrt{2}} ( A_{\lambda} + \frac{1}{\sqrt{2}} \kappa v_s) \nonumber \\
 &=& \frac{\sin2\beta}{2}( m_{H_u}^2 +m_{H_d}^2 + 2\mu_{\rm eff}^2 )+ \Delta_3 , \\
\Delta_3 &=&  \frac{\sin2\beta}{2}\frac{3 y_t^2}{16\pi^2} \left(
 \mathcal{F}(m_{\tilde{t}_1},Q) + \mathcal{F}(m_{\tilde{t}_2}, Q) - 2\mathcal{F}(m_t, Q) \right.  \nonumber \\  
&& \left.
 -\left(A_t^2+\mu_{\rm eff}^2- A_t\mu_{\rm eff}\frac{\tan\beta^2+1}{\tan\beta} \right)
   \frac{ \mathcal{F}(m_{\tilde{t}_1},Q) - \mathcal{F}(m_{\tilde{t}_2},Q)}{m_{\tilde{t}_2}^2-m_{\tilde{t}_1}^2} \right)  \nonumber \\
 &&     +\frac{\sin2\beta}{2} \frac{3 y_b^2}{16\pi^2}\left(
  \mathcal{F}(m_{\tilde{b}_1},Q) + \mathcal{F}(m_{\tilde{b}_2}, Q) - 2\mathcal{F}(m_b, Q) \right.  \nonumber \\  
&& \left.
 - \left( A_b^2+\mu_{\rm eff}^2-A_b\mu_{\rm eff}\frac{\tan\beta^2+1}{\tan\beta}  \right)
    \frac{\mathcal{F}(m_{\tilde{b}_1},Q) - \mathcal{F}(m_{\tilde{b}_2},Q)}{m_{\tilde{b}_2}^2-m_{\tilde{b}_1}^2} \right) \nonumber \\
 &&      + \frac{\sin2\beta}{2}\frac{ y_{\tau}^2}{16\pi^2} \left(
   \mathcal{F}(m_{\tilde{\tau}_1},Q) + \mathcal{F}(m_{\tilde{\tau}_2}, Q) - 2\mathcal{F}(m_\tau, Q) \right.  \nonumber \\  
&& \left.
 - \left( A_{\tau}^2+\mu_{\rm eff}^2-A_{\tau}\mu_{\rm eff} \frac{\tan\beta^2+1}{\tan\beta} \right)
    \frac{\mathcal{F}(m_{\tilde{\tau}_1},Q) - \mathcal{F}(m_{\tilde{\tau}_2},Q)}{m_{\tilde{\tau}_2}^2-m_{\tilde{\tau}_1}^2} \right) ,  
\end{eqnarray}
or alternatively
\begin{eqnarray}
(B\mu)_{\rm eff} &\equiv& \frac{\lambda v_s}{\sqrt{2}} ( A_{\lambda} + \frac{1}{\sqrt{2}} \kappa v_s) \nonumber \\
 &=&  \sin\beta\cos\beta \frac{\lambda^2 v^2}{2}
  - \sin\beta\cos\beta M_Z^2 - (m_{H_u}^2 -m_{H_d}^2) \frac{\tan\beta}{\tan\beta^2 - 1}
 + \Delta_4 , \\
\Delta_4
 &=&   - \frac{3 y_t^2}{16\pi^2}
 \left( \frac{\tan\beta}{\tan\beta^2 - 1} \left( \mathcal{F}(m_{\tilde{t}_1},Q) + \mathcal{F}(m_{\tilde{t}_2}, Q) - 2\mathcal{F}(m_t, Q) \right)  \right.  \nonumber \\  
&& \left.
 + \frac{ (\mu_{\rm eff}\tan\beta+A_t)(\mu_{\rm eff}- A_t\tan\beta)}{\tan\beta^2-1}\frac{ \mathcal{F}(m_{\tilde{t}_1},Q) - \mathcal{F}(m_{\tilde{t}_2},Q)}{m_{\tilde{t}_2}^2-m_{\tilde{t}_1}^2}  \right)  \nonumber \\
 && +\frac{3 y_b^2}{16\pi^2}
 \left( \frac{\tan\beta}{\tan\beta^2 - 1} \left( \mathcal{F}(m_{\tilde{b}_1},Q) + \mathcal{F}(m_{\tilde{b}_2}, Q) - 2\mathcal{F}(m_b, Q)\right)
  \right. \nonumber \\  && \left.
 + \frac{ (\mu_{\rm eff}+A_b\tan\beta)(\mu_{\rm eff}\tan\beta-A_b)}{\tan\beta^2 -1}  \frac{\mathcal{F}(m_{\tilde{b}_1},Q) - \mathcal{F}(m_{\tilde{b}_2},Q)}{m_{\tilde{b}_2}^2-m_{\tilde{b}_1}^2} \right) \nonumber \\
 && + \frac{y_{\tau}^2}{16\pi^2}
 \left( \frac{\tan\beta}{\tan\beta^2 - 1} \left(\mathcal{F}(m_{\tilde{\tau}_1},Q) + \mathcal{F}(m_{\tilde{\tau}_2}, Q) - 2\mathcal{F}(m_\tau, Q)\right)
  \right. \nonumber \\  && \left.
 + \frac{ (\mu_{\rm eff}+A_{\tau}\tan\beta)(\mu_{\rm eff}\tan\beta-A_{\tau})}{\tan\beta^2-1}\frac{\mathcal{F}(m_{\tilde{\tau}_1},Q) - \mathcal{F}(m_{\tilde{\tau}_2},Q)}{m_{\tilde{\tau}_2}^2-m_{\tilde{\tau}_1}^2} \right)  \nonumber \\
 && + \frac{1}{16\pi^2} \frac{\lambda \lambda_{N_i}}{2}
 \left(\mathcal{F}(m_{\tilde{N_i}_1},Q) - \mathcal{F}(m_{\tilde{N_i}_2}, Q) \right) .
\end{eqnarray}

The one loop corrected formula of Eq.~(\ref{eq:tadms2}) is
\begin{eqnarray}
m_S^2 &=& - \left( \frac{A_{\kappa} }{\sqrt{2}}\kappa v_s +\frac{ \lambda^2 v^2 }{2}
  - \kappa \lambda v^2\sin\beta\cos\beta + \kappa^2 v_s^2 \right) + \frac{A_{\lambda} \lambda v^2}{\sqrt{2}v_s}\sin\beta\cos\beta - \Delta_S ,\\
\Delta_S &=& -\frac{3 y_t^2}{16\pi^2}
 \frac{\mu_{\rm eff} v^2\cos\beta}{ v_s^2} (\mu_{\rm eff}\cos\beta - A_t\sin\beta )\frac{ \mathcal{F}(m_{\tilde{t}_1},Q) - \mathcal{F}(m_{\tilde{t}_2},Q)}{m_{\tilde{t}_2}^2-m_{\tilde{t}_1}^2}  \nonumber \\
&& - \frac{3 y_b^2 }{16\pi^2}\frac{ \mu_{\rm eff} v^2\sin\beta}{v_s^2}(\mu_{\rm eff}\sin\beta - A_b\cos\beta ) \frac{\mathcal{F}(m_{\tilde{b}_1},Q) - \mathcal{F}(m_{\tilde{b}_2},Q)}{m_{\tilde{b}_2}^2-m_{\tilde{b}_1}^2}  \nonumber \\
&& - \frac{ y_{\tau}^2}{16\pi^2}\frac{\mu_{\rm eff} v^2\sin\beta}{v_s^2}(\mu_{\rm eff}\sin\beta  - A_{\tau} \cos\beta ) \frac{ \mathcal{F}(m_{\tilde{\tau}_1},Q) - \mathcal{F}(m_{\tilde{\tau}_2},Q) }{m_{\tilde{\tau}_2}^2-m_{\tilde{\tau}_1}^2}   \nonumber \\
 && + \frac{\lambda_{N_i}}{16\pi^2 v_s} 
 \left[ 2\lambda_{N_i} v_s \left( \mathcal{F}(m_{\tilde{N_i}_1},Q) + \mathcal{F}(m_{\tilde{N_i}_2}, Q) - 2\mathcal{F}(m_{N_i}, Q) \right)  \right.  \nonumber \\
 && \left. -(\kappa v_s + \frac{1}{\sqrt{2}}A_{\lambda_{N_i}})\left(\mathcal{F}(m_{\tilde{N_i}_1},Q) - \mathcal{F}(m_{\tilde{N_i}_2}, Q) \right) \right] .
\end{eqnarray}

\section{Two-loop $\beta$ function for the singlet Higgs soft mass}
\label{app:twoloop}

We include here the two-loop $\beta$ function for $m_S^2$ with all Yukawa and trilinear couplings being complex: 
\begin{eqnarray}
{
	\beta_{m_S^2}^{(2)}}  &=  &
{
	-\frac{4}{5} (-3 g_1^2 |T_{\lambda}|^2 -15 g_2^2 |T_{\lambda}|^2 +120 m_S^2 \kappa^{2} \kappa^{* 2} +20 (m_{H_d}^2 + m_{H_u}^2 + m_S^2)\lambda^{2} \lambda^{* 2} +3 g_1^2 M_1 \lambda T_{\lambda}^* }\nonumber \\ 
&&
{ +15 g_2^2 M_2 \lambda T_{\lambda}^* +15 |T_{\lambda}|^2 \mbox{Tr}({y_d  y_d^{\dagger}}) +5 |T_{\lambda}|^2 \mbox{Tr}({y_e  y_e^{\dagger}}) +15 |T_{\lambda}|^2 \mbox{Tr}({y_u  y_u^{\dagger}})} \nonumber \\ 
&&
{+5 |T_{\lambda}|^2 \mbox{Tr}({y_\nu  y_{\nu}^{\dagger}}) +20 |T_{\kappa}|^2 \mbox{Tr}({\lambda_N  \lambda_N^*}) +15 \lambda T_{\lambda}^* \mbox{Tr}({y_d^{\dagger}  T_d}) +5 \lambda T_{\lambda}^* \mbox{Tr}({y_e^{\dagger}  T_e})} \nonumber \\ 
&&
{+15 \lambda T_{\lambda}^* \mbox{Tr}({y_u^{\dagger}  T_u}) +5 \lambda T_{\lambda}^* \mbox{Tr}({y_{\nu}^{\dagger}  T_{y_\nu}}) +20 \kappa T_{\kappa}^* \mbox{Tr}({\lambda_N^*  T_{\lambda_N}})} \nonumber \\ 
&&
{+\lambda^* (-3 g_1^2 m_{H_d}^2 \lambda -15 g_2^2 m_{H_d}^2 \lambda -3 g_1^2 m_{H_u}^2 \lambda -15 g_2^2 m_{H_u}^2 \lambda -3 g_1^2 m_S^2 \lambda -15 g_2^2 m_S^2 \lambda }\nonumber \\ 
&&
{+20 \lambda |T_{\kappa}|^2 +40 \lambda |T_{\lambda}|^2 +20 \kappa T_{\kappa}^* T_{\lambda} +3 g_1^2 M_1^* (-2 M_1 \lambda  + T_{\lambda})+15 g_2^2 M_2^* (-2 M_2 \lambda  + T_{\lambda})}\nonumber \\ 
&&
{+30 m_{H_d}^2 \lambda \mbox{Tr}({y_d  y_d^{\dagger}}) +15 m_{H_u}^2 \lambda \mbox{Tr}({y_d  y_d^{\dagger}}) +15 m_S^2 \lambda \mbox{Tr}({y_d  y_d^{\dagger}}) +10 m_{H_d}^2 \lambda \mbox{Tr}({y_e  y_e^{\dagger}}) \nonumber} \\ 
&&
{+5 m_{H_u}^2 \lambda \mbox{Tr}({y_e  y_e^{\dagger}}) +5 m_S^2 \lambda \mbox{Tr}({y_e  y_e^{\dagger}}) +15 m_{H_d}^2 \lambda \mbox{Tr}({y_u  y_u^{\dagger}}) +30 m_{H_u}^2 \lambda \mbox{Tr}({y_u  y_u^{\dagger}}) \nonumber} \\ 
&&
{+15 m_S^2 \lambda \mbox{Tr}({y_u  y_u^{\dagger}}) +5 m_{H_d}^2 \lambda \mbox{Tr}({y_\nu  y_{\nu}^{\dagger}}) +10 m_{H_u}^2 \lambda \mbox{Tr}({y_\nu  y_{\nu}^{\dagger}}) +5 m_S^2 \lambda \mbox{Tr}({y_\nu  y_{\nu}^{\dagger}}) \nonumber }\\ 
&&
{+15 T_{\lambda} \mbox{Tr}({T_d^*  y_d^{T}}) +15 \lambda \mbox{Tr}({T_d^*  T_{d}^{T}}) +5 T_{\lambda} \mbox{Tr}({T_e^*  y_e^{T}}) +5 \lambda \mbox{Tr}({T_e^*  T_{e}^{T}}) +15 T_{\lambda} \mbox{Tr}({T_u^*  y_u^{T}}) \nonumber }\\ 
&&
{+15 \lambda \mbox{Tr}({T_u^*  T_{u}^{T}}) +5 T_{\lambda} \mbox{Tr}({T_{{y \nu}^*}  y_{\nu}^{T}}) +5 \lambda \mbox{Tr}({T_{{y_ \nu}}^*  T_{y_\nu}^{T}}) +15 \lambda \mbox{Tr}({m_d^2  y_d  y_d^{\dagger}}) +5 \lambda \mbox{Tr}({m_e^2  y_e  y_e^{\dagger}}) \nonumber }\\ 
&&
{+5 \lambda \mbox{Tr}({m_l^2  y_e^{\dagger}  y_e}) +5 \lambda \mbox{Tr}({m_l^2  y_{\nu}^{\dagger}  y_\nu}) +15 \lambda \mbox{Tr}({m_q^2  y_d^{\dagger}  y_d}) +15 \lambda \mbox{Tr}({m_q^2  y_u^{\dagger}  y_u}) \nonumber }\\ 
&&
{+15 \lambda \mbox{Tr}({m_u^2  y_u  y_u^{\dagger}}) +5 \lambda \mbox{Tr}({m_{\tilde{N}}^2  y_\nu  y_{\nu}^{\dagger}}) )\nonumber }\\ 
&&
{+20 \kappa^* ((4 m_S^2  + m_{H_d}^2 + m_{H_u}^2)\kappa |\lambda|^2 +4 \kappa |T_{\kappa}|^2 +\kappa |T_{\lambda}|^2 +\lambda T_{\lambda}^* T_{\kappa} +4 m_S^2 \kappa \mbox{Tr}({\lambda_N  \lambda_N^*}) +T_{\kappa} \mbox{Tr}({\lambda_N  T_{{\lambda N}^*}}) \nonumber }\\ 
&&
{+\kappa \mbox{Tr}({T_{{\lambda N}^*}  T_{\lambda_N}}) +2 \kappa \mbox{Tr}({m_{\tilde{N}}^2  \lambda_N  \lambda_N^*}) )\nonumber }\\ 
&&
{+20 m_{H_u}^2 \mbox{Tr}({y_\nu  y_{\nu}^{\dagger}  \lambda_N  \lambda_N^*}) +20 m_S^2 \mbox{Tr}({y_\nu  y_{\nu}^{\dagger}  \lambda_N  \lambda_N^*}) +20 \mbox{Tr}({y_\nu  y_{\nu}^{\dagger}  T_{\lambda_N}  T_{{\lambda N}^*}}) +80 m_S^2 \mbox{Tr}({\lambda_N  \lambda_N^*  \lambda_N  \lambda_N^*}) \nonumber} \\ 
&&
{+20 \mbox{Tr}({\lambda_N  \lambda_N^*  T_{y_\nu}  T_{y_\nu}^{\dagger}}) +80 \mbox{Tr}({\lambda_N  \lambda_N^*  T_{\lambda_N}  T_{{\lambda N}^*}}) +20 \mbox{Tr}({\lambda_N  T_{{\lambda N}^*}  T_{y_\nu}  y_{\nu}^{\dagger}}) +80 \mbox{Tr}({\lambda_N  T_{{\lambda N}^*}  T_{\lambda_N}  \lambda_N^*}) \nonumber }\\ 
&&
{+20 \mbox{Tr}({\lambda_N^*  T_{\lambda_N}  T_{{y \nu}^*}  y_{\nu}^{T}}) +20 \mbox{Tr}({m_l^2  y_{\nu}^{\dagger}  \lambda_N  \lambda_N^*  y_\nu}) +20 \mbox{Tr}({m_{\tilde{N}}^2  y_\nu  y_{\nu}^{\dagger}  \lambda_N  \lambda_N^*}) \nonumber }\\ 
&&
{+20 \mbox{Tr}({m_{\tilde{N}}^2  \lambda_N  \lambda_N^*  y_\nu  y_{\nu}^{\dagger}}) +160 \mbox{Tr}({m_{\tilde{N}}^2  \lambda_N  \lambda_N^*  \lambda_N  \lambda_N^*}) +20 \mbox{Tr}({y_\nu  y_{\nu}^{\dagger}  \lambda_N  m_{\tilde{N}}^{2 *}  \lambda_N^*}) ), }
\label{eq:twoloop}
\end{eqnarray}
where $T_{i}$ stands for the trilinear parameter $A_i$ times the corresponding coupling $i$, where $i= y_i, \lambda, \kappa, \lambda_N$.


\small

\bibliographystyle{JHEP}
\bibliography{newlib}
\end{document}